\documentclass[superscriptaddress,aps,prb,twocolumn,floatfix]{revtex4-1}
\usepackage{amssymb, amsmath}
\usepackage{graphicx}
\usepackage{multirow}
\usepackage{color}

\usepackage[colorlinks,bookmarks=false,citecolor=darkblue,linkcolor=red,urlcolor=blue]{hyperref}
\definecolor{darkblue}{rgb}{0,0.02,0.45}
\newcommand{\fcs}[0]{FeCr$_2$S$_4$}

\usepackage{bm}
\usepackage{nicefrac}
\usepackage{hyperref}
\usepackage{booktabs}
\usepackage{multirow}

\begin{document}

\title{Magneto-electric properties and low-energy excitations of multiferroic FeCr$_2$S$_4$
}
\date{\today}

\author{A.~Strinic}
\author{S.~Reschke}
\affiliation{Experimentalphysik V, Center for Electronic
Correlations and Magnetism, Institute for Physics, Augsburg
University, D-86135 Augsburg, Germany}


\author{K.~V.~Vasin}
\affiliation{Institute for Physics, Kazan (Volga region) Federal University, 420008 Kazan, Russia}

\author{M.~Schmidt}
\affiliation{Experimentalphysik V, Center for Electronic
Correlations and Magnetism, Institute for Physics, Augsburg
University, D-86135 Augsburg, Germany}

\author{A.~Loidl}
\affiliation{Experimentalphysik V, Center for Electronic
Correlations and Magnetism, Institute for Physics, Augsburg
University, D-86135 Augsburg, Germany}

\author{V.~Tsurkan}
\affiliation{Experimentalphysik V, Center for Electronic
Correlations and Magnetism, Institute for Physics, Augsburg
University, D-86135 Augsburg, Germany} \affiliation{Institute of
Applied Physics, MD-2028~Chi\c{s}in\u{a}u, Republic of Moldova}

\author{M.~V.~Eremin}
\email{eremikhail@yandex.com}
\affiliation{Institute for Physics, Kazan (Volga region) Federal University, 420008 Kazan, Russia}

\author{J.~Deisenhofer}
\email{joachim.deisenhofer@physik.uni-augsburg.de}
\affiliation{Experimentalphysik V, Center for Electronic
Correlations and Magnetism, Institute for Physics, Augsburg
University, D-86135 Augsburg, Germany}

\date{\today}

\begin{abstract}
We report on the low-frequency optical excitations in the multiferroic ground state of polycrystalline \fcs in the frequency range 0.3-3~THz and their changes upon applying external magnetic fields up to 7~T.  In the ground state below the orbital-ordering temperature $T_{\mathrm{OO}}=9$~K we observe the appearance of several new modes. By applying the external magnetic field parallel and perpendicular to the propagation direction of the THz radiation, we can identify the strongest absorptions to be of predominantly electric-dipole origin.
We discuss these modes as the low-energy electronic excitations of the Fe$^{2+}$ ions ($3d^6$, $S\,=2$) in an tetrahedral S$^{2-}$ environment. The eigenfrequencies and relative intensities of these absorption lines are satisfactorily reproduced by our calculation assuming an effective exchange field of $12.8\;\mathrm{cm}^{-1}$ at the Fe$^{2+}$-ions sites. The direction of the exchange field is found to be slightly tilted out of the $ab$-plane. With our approach we can also describe previously reported results from Mössbauer studies and the order of magnitude of the electric polarisation induced by orbital and non-collinear spin ordering.


\end{abstract}

\maketitle


\section{Introduction}

\begin{figure}[b]
\includegraphics[clip,width=8cm]{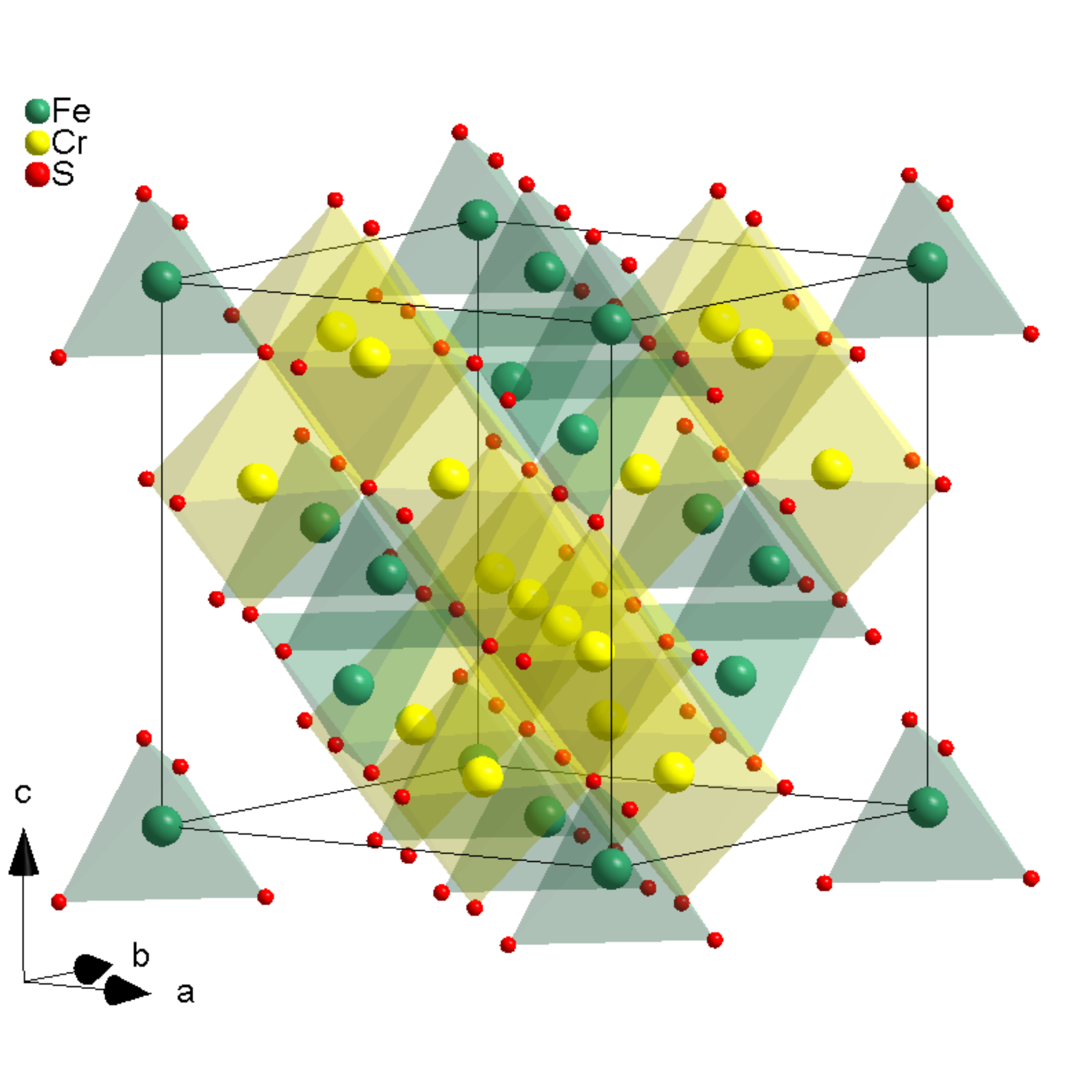}
\caption{\label{fig:structure} Structural unit cell of cubic \fcs with space group $Fd\bar{3}m$.}
\end{figure}

In condensed matter physics there are some materials, which were revisited again and again throughout several decades of research and always revealed exciting new properties.
One of these materials is the spinel FeCr$_2$S$_4$, which first came into the focus of research in the 1960s as a ferrimagnetic semiconductor with a $T_C = 170$~K \cite{Shirane1964,Lotgering1975}. Subsequently, the competing spin-orbit and electron-phonon interactions of the Jahn-Teller active Fe$^{2+}$-ions in tetrahedral environment were studied in the 1970-80s \cite{Englman1970,Spender1972,Brossard1979,Feiner1982a,Eibschutz1967,Hoy1968} and around the change of the millennium colossal magneto-resistance effects were reported \cite{Ramirez1997} and the magnetic structure was shown to be more complex than that of a simple collinear ferrimagnetic arrangement of the Cr$^{3+}$ ions ($3d^3$, $S=3/2$) on the octahedral sites and the Fe$^{2+}$ ions ($3d^6$, $S=2$) on the tetrahedral ones (see Fig.~\ref{fig:structure} for the cubic crystal structure with space group $Fd\bar{3}m$) \cite{Tsurkan2001,Tsurkan2001a,Tsurkan2001b,Maurer2003,Mertinat2005,Shen2009,Tsurkan2010}. In particular, a non-collinear magnetic structure is realized below $T_M=60$~K \cite{Kalvius2010}. In addition, a giant magneto-optical Kerr rotation was reported for the Fe$^{2+}$ $d-d$ transitions in the mid-infrared frequency range \cite{Ohgushi2005}. More recently, it was recognised that \fcs belongs to the class of materials with a multiferroic ground state, because the emergence of a finite polarization was reported below the orbital ordering transition at $T_{\mathrm{OO}}=$9~K \cite{Bertinshaw2014,Lin2014}. The electric polarization was reported to consist of two different contributions $P_1$ and $P_2$, with the latter arising directly at the transition temperature $T_{\mathrm{OO}}$, while the former appears at lower temperatures of around 4~K \cite{Lin2014}: In contrast to $P_1$ the contribution $P_2$ depends strongly on an external magnetic field and, therefore, was assigned to originate from a non-collinear spin configuration in the ground state. The component $P_1$ might then be a direct consequence of the structural Jahn-Teller distortion related to the orbital ordering.

\begin{figure}[ht]
\includegraphics[clip,width=8cm]{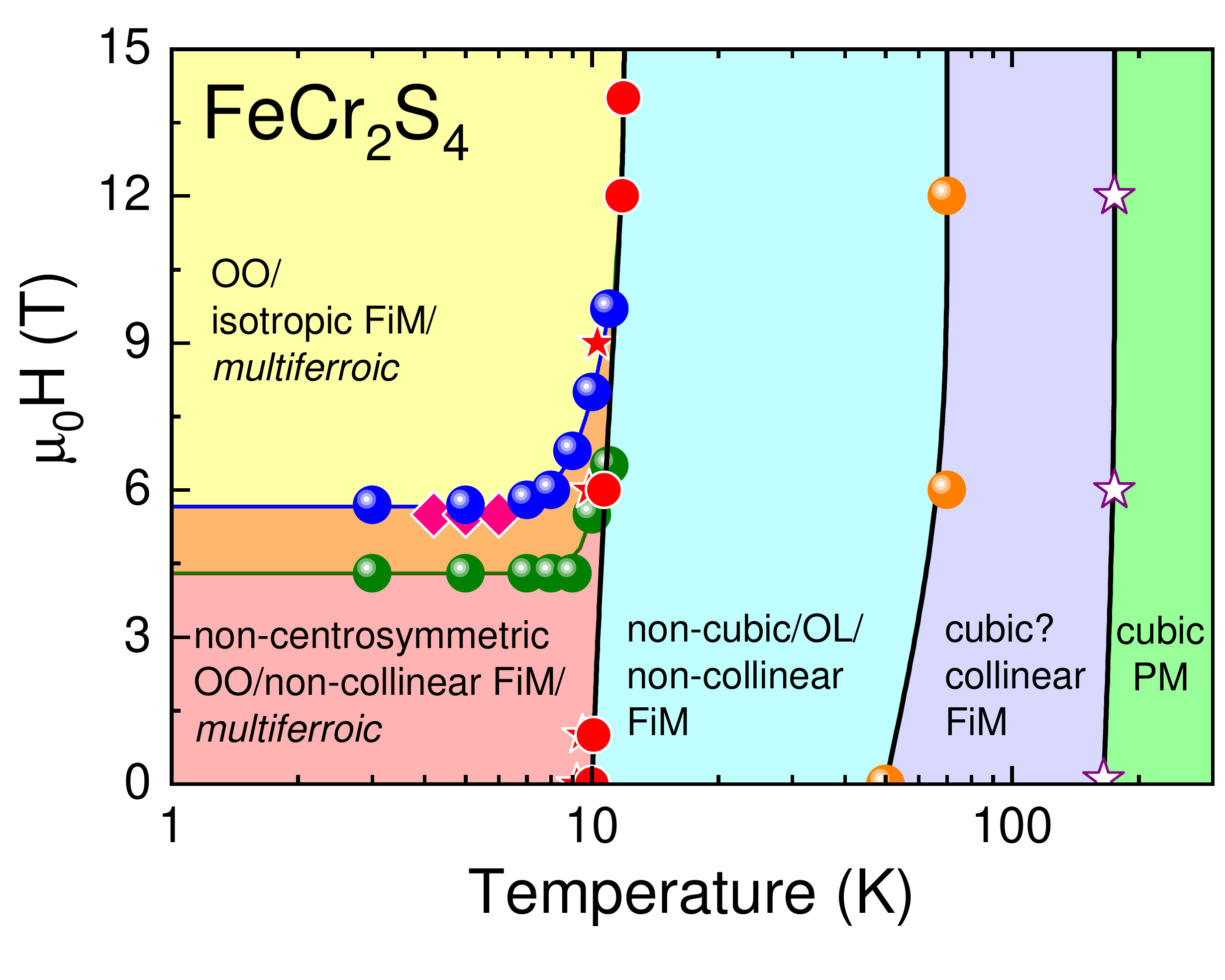}
\caption{\label{fig:Phasediagram} $H-T$ phase diagram of \fcs updated with respect to Ref.~\cite{Bertinshaw2014} by taking into account phonon-anomalies in zero-field reported in Ref.~\cite{Deisenhofer2019}.(FiM: ferrimagnet, PM: paramagnet, OO: orbital order, OL: orbital liquid)}
\end{figure}

While over the decades no direct evidence for clear deviations from cubic symmetry throughout the known phase transitions could be obtained \cite{Tsurkan2010,Bertinshaw2014}, some of us reported recently, that the transition temperatures $T_M$ and $T_{\mathrm{OO}}$ are accompanied by a splitting of infrared-active phonons and the emergence of new modes, confirming the expected lack of inversion symmetry in the multiferroic ground state \cite{Deisenhofer2019}. In Fig.~\ref{fig:Phasediagram} we show an updated $H-T$-phase diagram adapted from Ref.\cite{Bertinshaw2014}.

In this work we used THz-time domain spectroscopy to investigate possible optical magneto-electric effects in the multiferroic ground state and shed light on the mechanism involved in the formation of this state. We observed the emergence of several excitations in the multiferroic orbitally ordered state below $T_{\mathrm{OO}}$ and studied their dependence on external magnetic fields. In addition, we discuss a theoretical approach to model some of the observed excitations in terms of the low-energy electronic excitations of the Fe$^{2+}$ ions ($3d^6$, $S\,=2$) in an tetrahedral S$^{2-}$ environment.

\section{Experimental Details and Sample Properties}

We used polycrystalline samples with a high density of 3.85 g/cm$^3$ obtained by spark-plasma sintering
(SPS) technique.\cite{Tsurkan2010}. The density of the SPS sample is very close to the density of single crystals and, therefore, the measured absorption coefficients should not depend on the polycrystalline nature of the sample. Transmission measurements in the frequency range from 10-105\,cm$^{-1}$ were performed using THz-time-domain-spectroscopy  with a Toptica Tera-flash spectrometer and an Oxford Instruments cryomagnet in external magnetic fields up to 7\,T. The polycrystalline sample was polished to platelets with a thickness of 100~$\mu$m.

Due to the lack of single crystals which are large enough for long-wavelength optical measurements and at the same time exhibit the orbital ordering transition at 9~K, the measurement and light polarization configuration of the thin polycrystalline sample were restricted to the Faraday configuration with the wave vector $\mathbf{k}$ of the incoming linearly polarized THz pulse parallel to the external magnetic field $\mathbf{H}$ and the Voigt configuration with $\mathbf{k}\perp \mathbf{H}$. In the latter case the light polarization was varied to be either parallel $\mathbf{E}^{\omega}\parallel \mathbf{H}$ or perpendicular $\mathbf{E}^{\omega}\perp \mathbf{H}$ to the applied magnetic field.

\begin{figure}[t]
\includegraphics[clip,width=8.5cm]{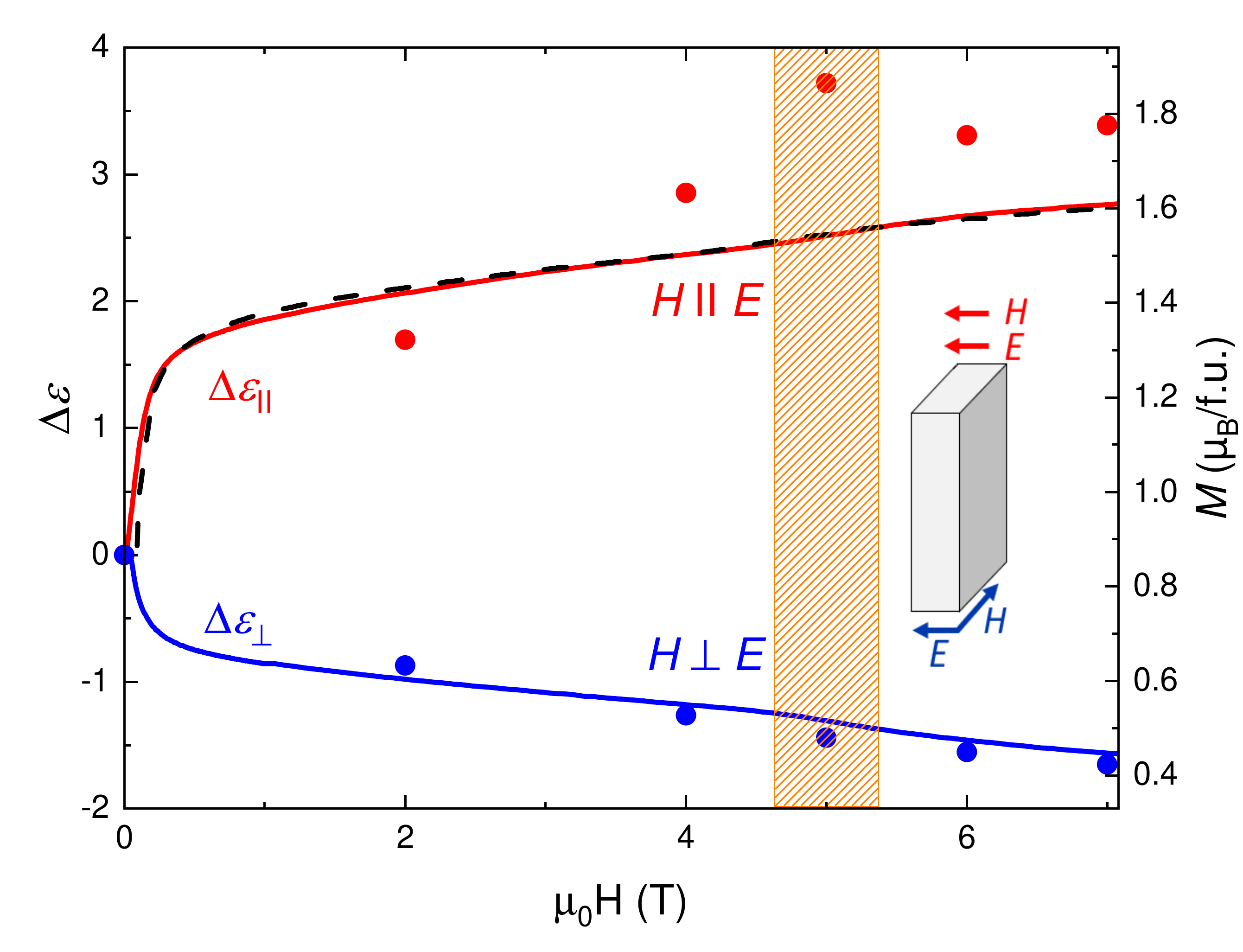}
\caption{\label{fig:deltaEps} Magnetic-field dependence of $\Delta\varepsilon^{\prime}$ defined by Eq.~\ref{Eq:epsilonstatic} with the probing microwave electric field parallel ($\Delta\varepsilon^{\prime}_\parallel$, red solid line) or perpendicular ($\Delta\varepsilon^{\prime}_\perp$, blue solid line) to the external magnetic field $H$ and the magnetization $M$ (black dashed line) taken from Ref.~\onlinecite{Bertinshaw2014}. The solid symbols represent corresponding estimates obtained by Eq.~\ref{eq:deltaeps+C} using the THz spectra as described in the text.}
\end{figure}

Before discussing our THz results in magnetic fields we want to recall the magnetic-field dependence of the magnetization and the dielectric constant of the polycrystalline sample. The magnetic-field dependence of the real-part of the dielectric constant at frequencies in the kHz-range was reported in Refs.\;\onlinecite{Lin2014,Bertinshaw2014} and used as an estimate of the static magnetic-field dependent value $\varepsilon^{\prime}_H(\omega\approx 0)$.
In Fig.~\ref{fig:deltaEps} we reproduce the change of $\varepsilon^{\prime}_H(\omega\approx 0)$ in an applied magnetic field at 5~K by showing the difference
\begin{equation} \label{Eq:epsilonstatic}
 \Delta \varepsilon^{\prime}(H)= \varepsilon^{\prime}_H(\omega\approx 0)-\varepsilon^{\prime}_{H=0}(\omega\approx 0)
\end{equation}
for the two cases that the external magnetic field was applied parallel and perpendicular to the electric field, denoted as $\Delta \varepsilon^{\prime}_\parallel$ and $\Delta\varepsilon^{\prime}_\perp$, respectively.
The value of $\varepsilon^{\prime}_{H=0}(\omega\approx 0)$ used here was determined as 21.4 in Ref.~\cite{Bertinshaw2014} in good agreement with 22.5 reported in Ref.~\cite{Lin2014}.

It was shown in  Ref.\onlinecite{Bertinshaw2014} that $\Delta \varepsilon^{\prime}_\parallel$ increases with respect to $H=0$ and $\Delta \varepsilon^{\prime}_\perp$ decreases with increasing magnetic field. In addition, it was reported that the magnetic-field dependence of the magnetization can be scaled on top of $\Delta \varepsilon^{\prime}_\parallel$ suggesting that the magnetic-field induced spin and magnetic domain reorientation is also responsible for the field-dependence of $\Delta \varepsilon(H)$.

\begin{figure*}[ht]
\includegraphics[keepaspectratio,width=17cm]{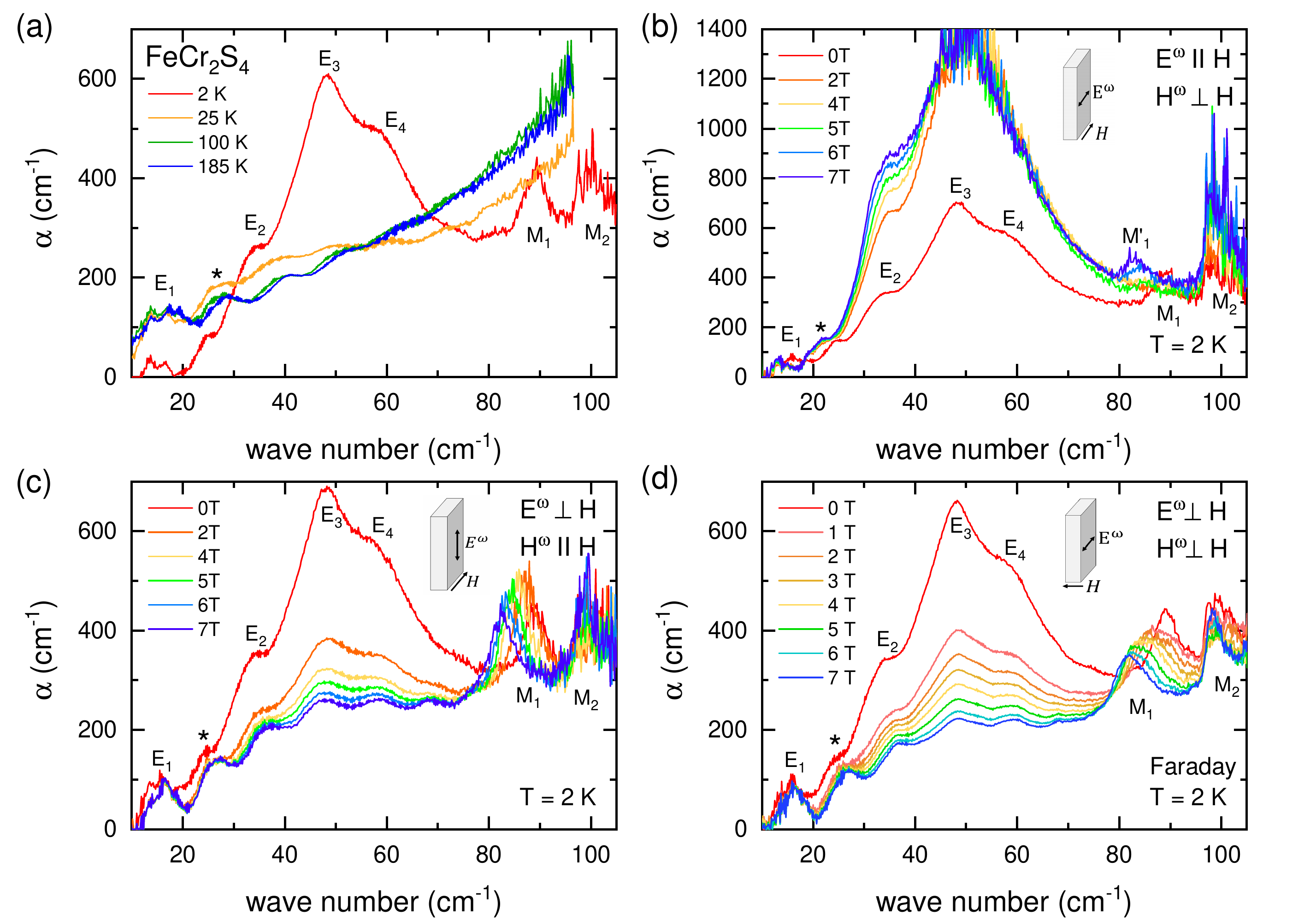}
\caption{\label{fig:Voigtspectra} Absorption spectra for (a)  different temperatures upon zero-field cooling, (b) at 2~K  for increasing magnetic fields in Voigt configuration with $E^\omega\parallel H, H^\omega\perp H$ and (c) for $E^\omega\perp H, H^\omega\parallel H$. (d) Absorption spectra
at 2~K  for increasing magnetic fields in Faraday configuration with $E^\omega\perp H, H^\omega\perp H$.}
\end{figure*}

Looking at the $M-H$-dependence of the polycrystalline sample, it is important to note that in single-crystal \fcs \,there is a clear magnetocrystalline anisotropy with the easy axis coinciding with the $\langle 001 \rangle$ direction, while the  $\langle 110 \rangle$ and $\langle 111 \rangle$ directions are considered to be hard axes \cite{Tsurkan2001, Fritsch2003}. The data of the polycrystal represent a statistical average, where nearly all domains become coaligned in fields below 1~T and the increase towards higher fields is due to the successive alignment of the magnetization in the domains where the external magnetic field has to overcome the magneto-crystalline anisotropy. It is anticipated that the reported increase of the magnetization in the range of $4.5~\mathrm{T}<H<5.5 ~\mathrm{T}$ (indicated by the shaded area) is related to the critical fields necessary to overcome the magneto-crystalline anisotropy in \fcs \cite{Felea2014}.

\section{Experimental Results and Discussion}

\subsection{Temperature dependence}

In Figure \ref{fig:Voigtspectra}(a) we show the absorption coefficient of \fcs\,as a function of temperature.  The temperature dependence upon cooling the sample in Figure \ref{fig:Voigtspectra}(a) reflects the sequence of the different magnetic phases shown in the phase diagram in Fig.~\ref{fig:Phasediagram}(b). The monotoneous increase of the absorption coefficient with frequency at 185~K and 100~K representing the paramagnetic and collinear ferrimagnetic phases, respectively, is attributed to the lowest-lying infrared-active phonon at about 125~cm$^{-1}$, which narrows and undergoes a blueshift with decreasing temperatures \cite{Wakamura1989,Rudolf2005,Deisenhofer2019}. As a result the accessible frequency range for transmission measurements within our experimental sensitivity increases with decreasing temperatures and in the non-collinear magnetic phase for $9~\mathrm{K}<T<60~\mathrm{K}$ spectral weight is shifted to the THz-range indicated by
a broad absorption continuum, which appears below 85cm$^{-1}$ in the spectrum at 25~K. This absorption continuum increases in intensity on approaching the orbital ordering transition at $T_{OO}=9$~K and develops into a strong absorption band within the multiferroic ground state, where in the spectrum at 2~K a minimal set of four distinct modes  $E_1-E_4$ may be distinguished by the shoulders and maxima of the absorption band as indicated in in Fig.~\ref{fig:Voigtspectra}(a). In addition, two more modes, $M_1$ and $M_2$, appear in the polar ground state at 90~cm$^{-1}$ and 100~cm$^{-1}$, respectively.

Note that for $T>T_{\mathrm{OO}}$ there is a clear periodic modulation of the spectra, which can be explained as a result of internal reflections in the platelet-shaped sample. For example, the maxima in the spectra at 100~K are separated by 11~cm$^{-1}$ in agreement with a thickness of 100~$\mu$m and a refractive index of $n=4.5$. We  interpret  the  feature between $E_1$ and $E_2$ (marked by  an asterisk) at about  27~cm$^{-1}$ as  a remainder of this Fabry-Perot interference pattern. Consequently, it should mainly be affected by the magnetic field due to a change of the refractive index and the changing contributions of $E_1-E_4$. The field dependence shown in Fig.~\ref{fig:e1intensity} supports this assignment.

\subsection{Field dependence}

In Figs.~\ref{fig:Voigtspectra}(b) and (c) we show the evolution of the spectra measured at 2~K in increasing magnetic fields in Voigt configuration with
$E^\omega\parallel H, H^\omega\perp H$ and $E^\omega\perp H, H^\omega\parallel H$, respectively, and in Faraday configuration with $E^\omega\perp H, H^\omega\perp H$, in Fig.~\ref{fig:Voigtspectra}(d):

In the configuration $E^\omega\parallel H, H^\omega\perp H$
in Fig.~\ref{fig:Voigtspectra}(b), there is a huge increase of the intensity of the $E$ absorption band in the frequency range 20-80~cm$^{-1}$, while the maxima and shoulders of $E_2-E_3$ do not exhibit a discernible shift with increasing fields. For $E^\omega\perp H, H^\omega\parallel H$ in Fig.~\ref{fig:Voigtspectra}(c), again, the most significant changes take place in the frequency range 20-80~cm$^{-1}$ and between 0T and 2T, where the excitations $E_2-E_4$ become strongly suppressed with increasing magnetic fields. Note that the absorption spectrum at 7~T in this frequency range clearly resembles the zero-field cooled spectrum at 25~K shown in Fig.~\ref{fig:Voigtspectra}(a). In Faraday configuration with $E^\omega\perp H, H^\omega\perp H$ the changes of $E_2-E_4$ with increasing fields are similar to measurements in Voigt configuration $E^\omega\perp H, H^\omega\parallel H$. This indicates that the orientation of the THz-electric field $E^\omega$ with respect to the static magnetic field $H$ is the decisive factor for the huge changes in intensity of the $E_2-E_4$ absorptions.

\begin{figure}[h]
\includegraphics[clip,width=8cm]{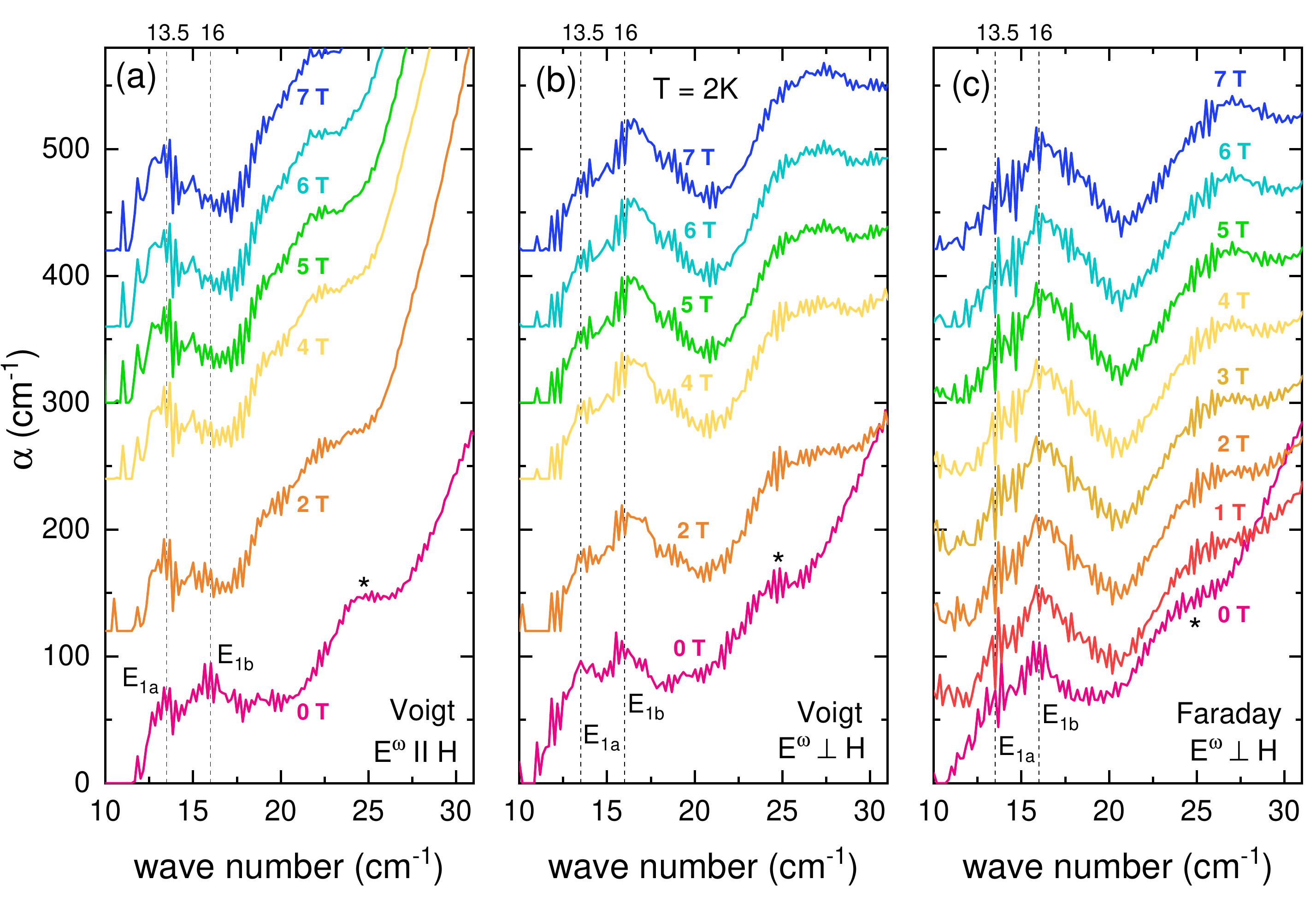}
\caption{\label{fig:e1intensity} Magnetic field dependence of mode $E_1$ at 2~K  for (a) with $E^\omega\parallel H, H^\omega\perp H$, (b) for $E^\omega\perp H, H^\omega\parallel H$, and (c) for $E^\omega\perp H$, $H^\omega\perp H$. The spectra were shifted by a vertical offset with respect to the zero-field curve for clarity. Dashed lines indicate absorption peaks in zero-field as described in the text.}
\end{figure}

In contrast, the behavior of $E_1$ is more complex and shown in detail in Fig.~\ref{fig:e1intensity} for $T=2$~K: For $E^\omega\parallel H, H^\omega\perp H$  (Fig.~\ref{fig:e1intensity}(a)) we first identify two absorption peaks $E_{1a}$ and $E_{1b}$ at 13.5 and 16.0~cm$^{-1}$ (dashed lines) in the zero-field spectrum, respectively. With increasing field in the range 0-2~T the intensity of $E_{1b}$ decreases and remains at a lower level for higher fields.
For $E^\omega\perp H, H^\omega\parallel H$  (Fig.~\ref{fig:e1intensity}(b)) both absorption peaks are also visible and an increase in intensity for $E_{1b}$ in the range 0-2~T can be anticipated.
In Faraday configuration the  spectrum exhibits a single maximum at the eigenfrequency of $E_{1b}$ and does not change significantly with increasing magnetic field. As mode $E_{1a}$ does not discern itself by a clear peak, it is impossible to judge about its presence in this case.
Although changes in this frequency range are evident, we want to point out that the strong changes of the absorptions $E_2-E_4$ clearly influence the changes in the region of $E_1$.

\begin{figure}[ht]
\includegraphics[clip,width=9cm]{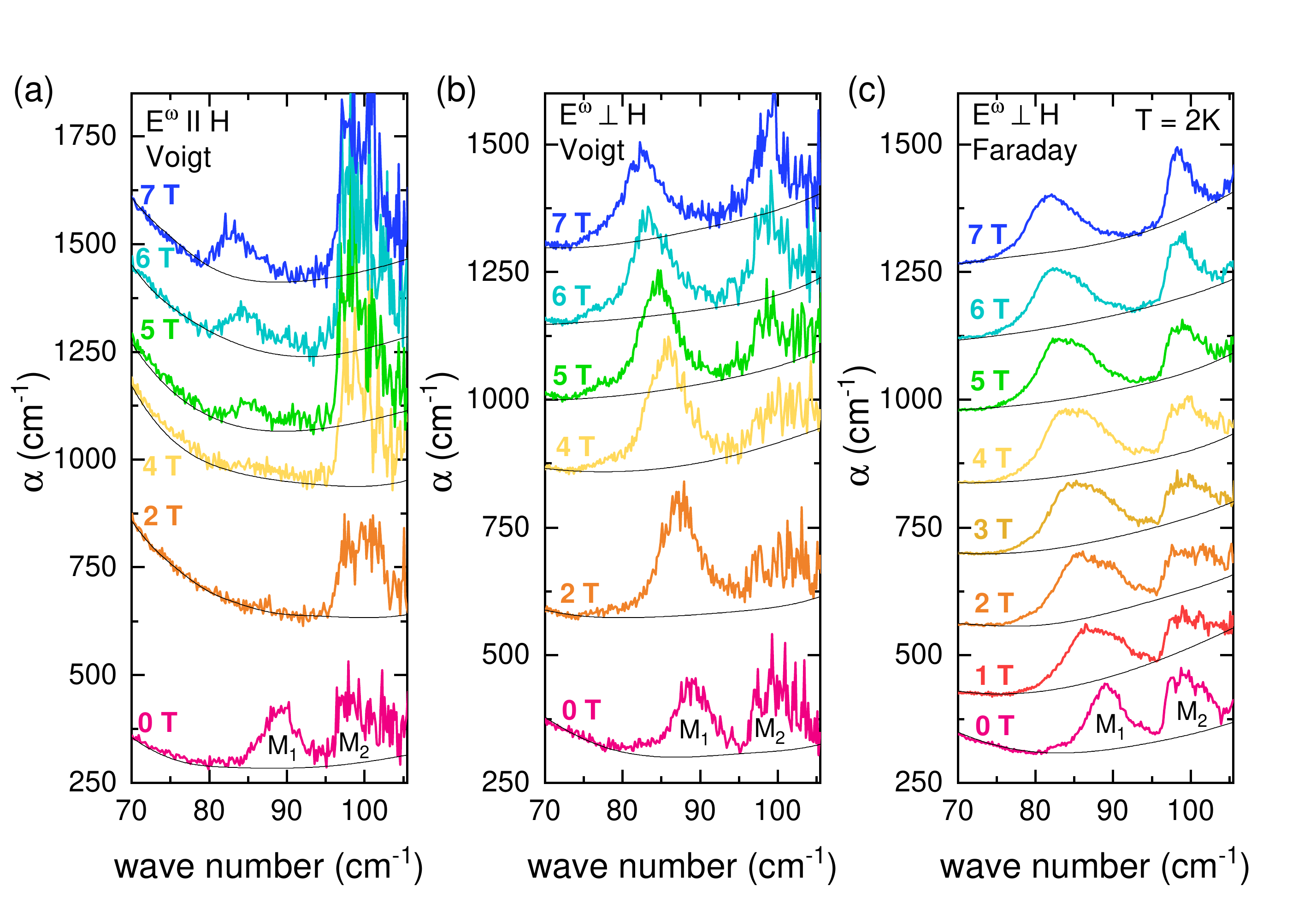}
\caption{\label{fig:ZoomIn} Magnetic field dependence of modes $M_1$ and $M_2$  at 2~K  for (a) with $E^\omega\parallel H, H^\omega\perp H$, (b) for $E^\omega\perp H, H^\omega\parallel H$, and (c) for $E^\omega\perp H, H^\omega\perp H$. The spectra were shifted by a vertical offset with respect to the zero-field curve for clarity. Solid lines are fitting curves of the baseline as described in the text.}
\end{figure}

Interestingly, a similar behavior can be observed for $M_1$, and $M_2$ which is highlighted in Fig.~\ref{fig:ZoomIn} for all three configurations:
While $M_2$ shows an increase in absorption with increasing magnetic field $E^\omega\parallel H, H^\omega\perp H$ (Fig.~\ref{fig:ZoomIn}(a)), mode $M_1$ is not visible anymore at 2~T, but a somewhat weaker absorption feature emerges again for fields larger than 4~T and shifts to lower frequencies with further increasing fields. For the configurations $E^\omega\perp H, H^\omega\parallel H$ shown in Fig.~\ref{fig:ZoomIn}(b), there is first an increase in intensity of $M_1$ between 0~T and 2~T, and then a decrease in intensity for field larger than 2~T. There is a shift to lower frequencies with increasing fields. Intensity changes of $M_2$ in this configuration can not be tracked reliably due to the lower transmission in this frequency range. In Faraday configuration (see Fig.~\ref{fig:ZoomIn}(c)) the lineshapes of $M_1$ and $M_2$ are better resolved and the intensity seems to be slightly reduced in comparison to Fig.~\ref{fig:ZoomIn}(b). Again, M$_1$ gains intensity in the range 0-2~T and shifts to lower frequencies with increasing magnetic field. Significant intensity changes of $M_2$ are, however, still difficult to track.

Before we discuss the possible origin and optical activity of the observed modes, we want to point out that the most significant changes in magnetic field, the increase in intensity of the $E$-band and the disappearance of $M_1$ happen between 0T and 2T, coinciding with the field regime, where the majority of the magnetic and ferroelectric domains are getting aligned as discussed above for the magnetization $M$ and $\Delta \varepsilon^{\prime}_\parallel$ in Fig.~\ref{fig:deltaEps}. The reappearance of $M_1$ for $H>2 \mathrm{T}$ is tentatively assigned to the changes in the field regime $4.5~\mathrm{T}<H<5.5 ~\mathrm{T}$, which are associated with an overcoming of the magneto-crystalline anisotropy by the external magnetic field \cite{Bertinshaw2014}.

\subsection{Discussion}

The observed increase or decrease of the intensity of the E-band for the respective configurations $E^\omega\parallel H$ and $E^\omega\perp H$ and the agreement with the field variation of the static real part of the dielectric function $\Delta \varepsilon(H)$ shown in Fig.~\ref{fig:deltaEps} strongly suggests that the excitations are predominantly of electric-dipole origin. As a consequence we assume that the $E$-band excitations can be assigned to the  imaginary part of the dielectric constant $\varepsilon^{\prime\prime}(\omega)$, which is related via the Kramers-Kronig transformation to the static real part of the dielectric constant $\varepsilon^{\prime}(\omega)$:

\begin{equation}
  \varepsilon^{\prime}_H(0)-1=\frac{2}{\pi}\int_{0}^{\infty}\frac{\varepsilon^{\prime\prime}_H(\omega)}{\omega}d\omega
\end{equation}

To check the validity of this assumption, we evaluate the experimental data by converting the time-domain spectra in terms of the dielectric function and integrate from $\omega_1$ = 21~cm$^{-1}$ to $\omega_2$ = 76~cm$^{-1}$, thus concentrating on the excitations $E_2$-$E_4$, which show the most prominent changes. Hence, we compare the obtained dynamical quantity
\begin{equation}\label{eq:deltaeps+C}
  \delta\varepsilon^{\prime}(H)=\frac{2}{\pi}\int_{\omega_1}^{\omega_2}\left(
  \frac{\varepsilon^{\prime\prime}_H(\omega)}{\omega}-\frac{\varepsilon^{\prime\prime}_{H=0}(\omega)}{\omega}\right)d\omega + C
\end{equation}
with $\Delta \varepsilon^{\prime}(H)$  in Fig.~\ref{fig:deltaEps}. Thereby, we assume that the dependence of $\varepsilon_H^{\prime\prime}(\omega)$ on the magnetic field outside this frequency range can be approximated by a constant contribution $C$, which can be different for the two configurations $E^\omega\parallel H$ and $E^\omega\perp H$. Using the values $C_\parallel=-4.8$ and $C_\perp=+1.9$ for $H>0$ the two quantities scale nicely, in particular in the case $E^\omega\perp H$. In the case $E^\omega\parallel H$ the maxima of the strong absorption bands could not be resolved anymore, resulting in a larger uncertainty due to a possible underestimation of the dielectric strength of the $E$-band. This should, in principle, produce lower values in comparison to the microwave data, which is not the case. Indeed, we even had to used the negative value $C_\parallel=-4.8$ to even further reduce the integrated dielectric strength in an attempt to show that both quantities follow a similar trend. As it will be discussed below in section IV, the underlying electronic levels of the $E$-band excitations allow also for magnetic-dipole contributions and interference effects, which strongly affect the intensity of the strongest $E$-band excitations and shows the limits of this comparison.
However, we think that this is clear evidence for a predominant electric-dipole origin of the excitations $E_2$-$E_4$. Their eigenfrequencies in zero field are denoted by $\omega_0$ in the upper part of Tab.~\ref{tab:selectionrules} together with their occurrence or suppression in fields of 7~T. Assuming that the primary effect of the external magnetic field is to align magnetic domains and overcome the magnetic anisotropy, the increase of the absorption of the $E$-band should be a result of aligning the  electric dipoles responding to the electric-field component of the THz radiation. This is in agreement with the interpretation of the reported polarization component $P_2$ originating from the non-collinear spin configuration in the orbitally ordered state \cite{Lin2014}, with the corresponding ferroelectric domains strongly linked to the magnetic ones.

\begin{figure}[h]
\includegraphics[clip,width=9cm]{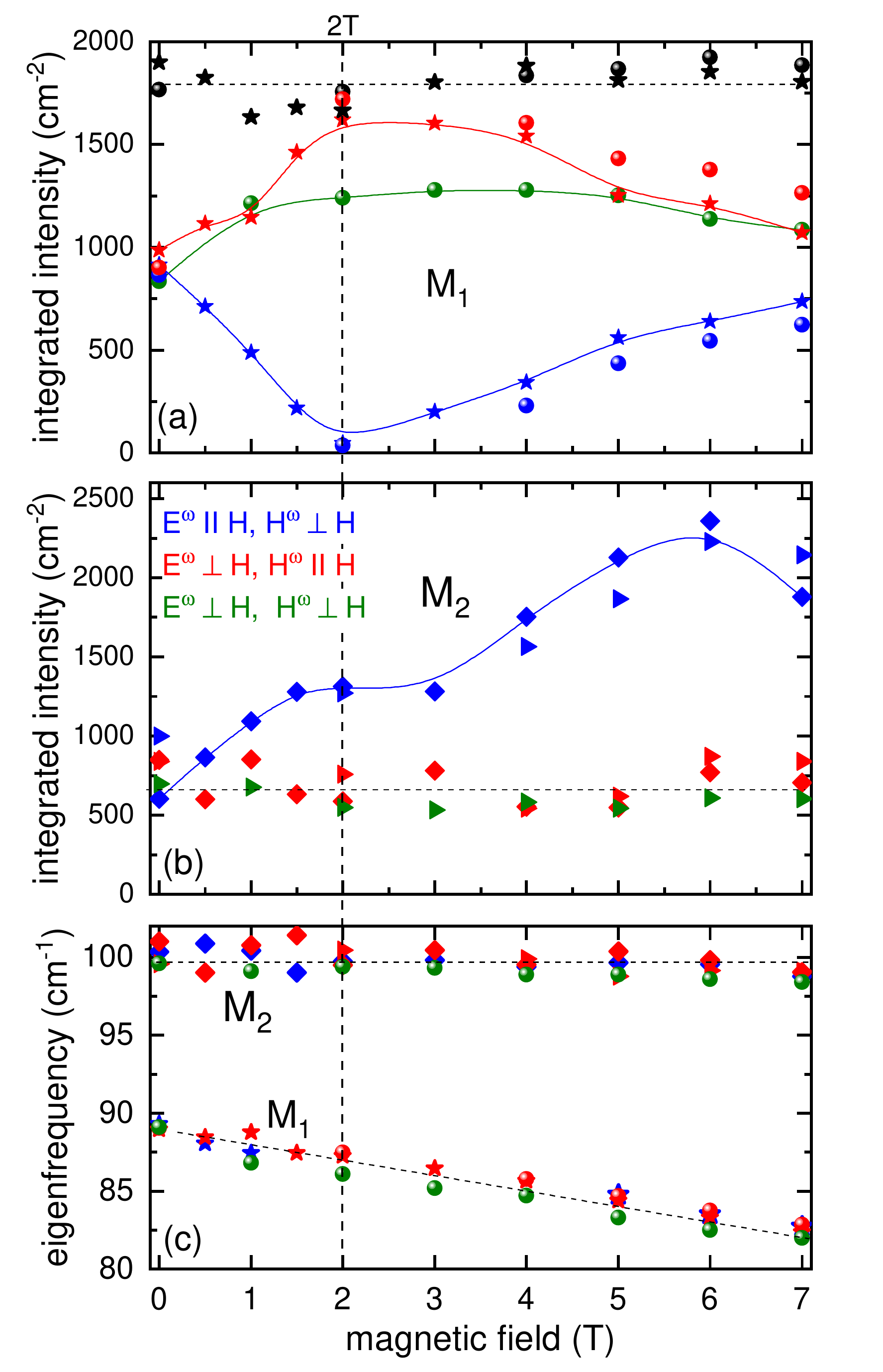}
\caption{\label{fig:m1m2intensity} Magnetic field dependence for the three configurations $E^\omega\parallel H, H^\omega\perp H$ (blue), $E^\omega\perp H, H^\omega\parallel H$ (red), and $E^\omega\perp H, H^\omega\perp H$ (green) of (a) the integrated absorption coefficient of mode $M_1$ at 2~K (spheres) and 5~K (stars) and the sum for the two Voigt configurations (black) and (b) the integrated absorption coefficient of mode $M_2$ at 2~K (diamonds) and 5~K (triangles). In (c) the eigenfrequency of the modes for the different polarization configurations at 2~K and 5~K are shown. Lines are to guide the eyes.}
\end{figure}

Similarly, the changes of $M_1$ and $M_2$ with increasing field in the three configurations can be understood in terms of the magnetic field dependence of the magnetization. In order to compare the relative changes of the intensity of $M_1$ and $M_2$, we used polynomial fits (shown as black lines in Fig.~\ref{fig:ZoomIn}) outside of the frequency region of $M_1$ and $M_2$ to subtract the underlying absorption background due to the lower-lying $E$-band and higher-lying phonon contributions and then integrated over the absorption peaks.
The results for $M_1$ are displayed in Fig.~\ref{fig:m1m2intensity}(a), where the values for 2~K and for 5~K (spectra not shown) are shown.

 \begin{table}[t]
\squeezetable
\begin{ruledtabular}
\centering\footnotesize
   \begin{tabular}{cccccc}
    \multirow{3}{*}{\textbf{mode}}
	& {\bfseries $\boldsymbol{\omega_0}$} &
	\textbf{Faraday at 7~T} &
	\multicolumn{2}{c}{\textbf{Voigt at 7~T}} &
	\multirow{3}{*}{\textbf{activity}} \\
    & $[\mathrm{cm}^{-1}]$ & $E^{\omega}\perp H$ & $E^{\omega}\perp H$
    & \multicolumn{1}{c}
     {$E^{\omega}\parallel H$}  &  \\
    & &$H^{\omega}\perp H$ & $H^{\omega}\parallel H$
    & \multicolumn{1}{c}
     {$H^{\omega}\perp H$} &  \\
    \midrule
    $E_2$ & 36 &  $\times$  &  $\times$ &    $\checkmark$
	 &  $E^{\omega}\parallel H$     \\
	 $E_3$ & 48 &$\times$  &  $\times$ &    $\checkmark$
	 &  $E^{\omega}\parallel H$ 	\\
	 $E_4$& 56 & $\times$  &  $\times$ &    $\checkmark$
	 &  $E^{\omega}\parallel H$   \\
    \midrule
    \midrule
     \multirow{3}{*}{\textbf{mode}}
	& {\bfseries $\boldsymbol{\omega_0}$} &
	\textbf{Faraday at 2~T} &
	\multicolumn{2}{c}{\textbf{Voigt at 2~T}} &
	\multirow{3}{*}{\textbf{activity}} \\
    & $[\mathrm{cm}^{-1}]$ & $E^{\omega}\perp H$ & $E^{\omega}\perp H$
    & \multicolumn{1}{c}
     {$E^{\omega}\parallel H$}  &  \\
    & &$H^{\omega}\perp H$ & $H^{\omega}\parallel H$
    & \multicolumn{1}{c}
     {$H^{\omega}\perp H$} &  \\
    \midrule
	 $E_{1a}$ & 13.5 &  n.d. & $\checkmark$  &  $\checkmark$
	 &   $E^{\omega}\perp H$,    \\
	  &  &    &     &
	 &  $E^{\omega}\parallel H$     \\
	  $E_{1b}$ & 16 &  $\checkmark$ & $\checkmark$  &  $\times$
	 &  $E^{\omega}\perp H$     \\
	
	 $M_1$ & 90 &  $\checkmark$  & $\checkmark$     &  $\times$
	 &  $E^{\omega}\perp H$     \\
	 $M_2$ & 100 &  $\checkmark$  & $\checkmark$    &  $\checkmark$  &  $E^{\omega}\perp H$,     \\
	  &  &    &     &
	 &  $E^{\omega}\parallel H$     \\
	
    \end{tabular}
    \caption{ \label{tab:selectionrules} Eigenfrequencies in zero field $\omega_0$ and observed optical activity for Voigt and Faraday configurations. Upper part: The  notation $\checkmark$
    and $\times$ indicates that the mode's intensity is strongly increased  or strongly suppressed in a magnetic field of 7~T, respectively.
    Lower part: The  notation $\checkmark$
    and $\times$ indicates that the mode is present or strongly suppressed in a magnetic field of 2~T, respectively. The notation n.d. signifies that the mode was not clearly descernible.}
\end{ruledtabular}
\end{table}

 Clearly, the magnetic field dependence reveals the suppression of $M_1$ in fields of 2~T for $E^\omega\parallel H, H^\omega\perp H$ and an increase for $E^\omega\perp H, H^\omega\parallel H$. For further increasing fields the integrated intensity again decreases for $E^\omega\perp H, H^\omega\parallel H$ and increases for $E^\omega\parallel H, H^\omega\perp H$.  Notably, plotting the sum of the intensities of $M_1$ for the two configurations can, however, be considered constant within the experimental uncertainty. This implies that domain reorientation in the applied magnetic field can induce a configuration in a field of about 2~T, which determines a selection rule for the optical activity of $M_1$, namely $E^\omega\parallel H, H^\omega\perp H$. For further increasing fields this metastable configuration is again lost, presumably due to the interplay of the magnetocrystalline anisotropy and the external magnetic field \cite{Bertinshaw2014}.

 The integrated intensity of $M_1$ in Faraday configuration $E^\omega\perp H, H^\omega\perp H$ follows a similar but somewhat less pronounced trend as for $E^\omega\perp H, H^\omega\parallel H$. Hence, we conclude that in the field range $0<H<3$~T mode $M_1$ is mainly active for $E^\omega\perp H$ as denoted in the lower part of Tab.~\ref{tab:selectionrules}.
 In the case of mode $M_2$ the integrated intensity shown in Fig.~\ref{fig:m1m2intensity}(b) has to considered with care, because in both Voigt configurations the maxima of the absorption peaks could not be resolved as discussed above. However, it seems that the intensities for $E^\omega\perp H, H^\omega\parallel H$ and $E^\omega\perp H, H^\omega\perp H$ are in agreement with each other and do not depend strongly on the applied magnetic field. For $E^\omega\parallel H, H^\omega\perp H$ the intensity of mode $M_2$ shows an increase with increasing magnetic field and an overall higher intensity than for the other configurations, suggesting the presence of an additional excitation mechanism for $M_2$. Unfortunately, the exact nature of $M_1$ and $M_2$ cannot be established based on the present data and measurements on single crystals will be necessary to determine, if they correspond to collective magneto-electric magnon modes.

 A similar ambiguity remains for the optical activity of excitation $E_{1a}$, while for $E_{1b}$ below 2~T the optical activity is assigned as $E^\omega\perp H$ in Tab.~\ref{tab:selectionrules}.
 Finally, the magnetic-field dependence of the eigenfrequencies of $M_1$ and $M_2$ are shown in Fig.~\ref{fig:m1m2intensity}(c), showing that $M_1$ exhibits a similar shift to lower frequency with increasing field for all configurations, while the eigenfrequency of $M_2$ remains approximately constant.

\section{Theoretical Model for the $E$-band Excitations}

In the following we will discuss a model, which describes the excitations between the lowest-lying electronic $d-d$- levels of the Fe$^{2+}$-ions in tetrahedral environment
\cite{Slack1966,Slack1969,Wittekoek1973,Mittelstaedt2015,Laurita2015,Mai2016}. We will discuss the effective single-ion Hamiltonian and compare the resulting absorption scheme with the experimentally observed excitations forming the $E$-band.

\subsection{Effective Hamiltonian for the low-lying Fe$^{2+}$ states in the presence of electric and magnetic fields}\label{sec:low-lying-model}

The Fe$^{2+}$-ions occupy the tetrahedral sites with electronic configuration $t_{2}^3 e^{3}$ in the ground state, while the Cr$^{3+}$-ions at the centre of S$^{2-}$ octahedra are in $t_{2g}^{3}$ configuration, which has no orbital degrees of freedom. In the \fcs unit cell there are two tetrahedral fragments Fe(1)S$_4$ and Fe(2)S$_4$ which are rotated relative to each other by $90^{\circ}$  around the $c$-axis.

It is well known that the level scheme of the low-lying $^5$E-states depends on the competition of spin-orbit coupling, the Jahn-Teller-effect, and, in the case of magnetically ordered systems, on the internal magnetic exchange fields at the Fe$^{2+}$-sites \cite{Vallin1970,Varret1972,Bonville1981a,Feiner1982}.

For the calculation of the Fe$^{2+}$-ions energy levels we used the effective Hamiltonian \cite{Ono1954,Fujiwara1974}
\begin{eqnarray}
\label{eq:Heff}
&&H_{eff} = -\xi\Big\{ [3 S_z^2 - S(S+1)]U_{\theta}+\frac{\sqrt{3}}{2}(S_{+}^2+S_{-}^2)U_{\varepsilon} \Big\} \nonumber\\
&&+ \zeta\Big( S_x S_z S_y + S_y S_z S_x + S_z S_x S_y + S_z S_y S_x +\nonumber\\
&& + S_x S_y S_z + S_y S_x S_z \Big)U_\alpha  + V\rho \Big(U_\theta \cos(\phi) + U_\varepsilon \sin(\phi)\Big)\nonumber\\
&&+  \sum  J_{Fe,j} \mathbf{S}  \langle \mathbf{S}_j\rangle + \Big(g_s-\frac{4\lambda}{\Delta}\Big)\mu_B \mathbf{B}\mathbf{S}- \frac{2\lambda\mu_B}{\Delta}\times\nonumber\\
&& \times \Big\{ (3 S_z B_z - \mathbf{B}\mathbf{S})U_\theta+\sqrt{3}(B_x S_x-B_y S_y)U_\varepsilon \Big\}.
\end{eqnarray}
The first two terms account for the spin-spin and spin-orbit interactions with $\xi = \rho_S + \lambda^2/\Delta+2\lambda^3/\Delta^2$ and $\zeta = \sqrt{3}\lambda^3/\Delta^2$. The parameter $\Delta$ denotes the crystal-field splitting between the exited ${}^5 T_{2}$ and ground ${}^5 E_{2}$ states, and $\lambda$ is the spin-orbit coupling constant. The Pauli-like matrices $U_\theta = |\varepsilon\rangle\langle{\varepsilon}| - |\theta\rangle\langle\theta|$ and $U_\varepsilon = |\varepsilon\rangle\langle\theta| + |\theta\rangle\langle\varepsilon|$, $U_\alpha = i (|\theta\rangle\langle\varepsilon|-|\varepsilon\rangle\langle\theta|)$  describe the E-orbital ground state within the orbital doublet states $|\theta\rangle$ and $|\varepsilon\rangle$. The third term takes into account a possible distortion of the FeS$_4^{2-}$ tetrahedra at low temperatures due to a linear Jahn-Teller coupling of the E-orbital states. The last three terms in Eq.\;\eqref{eq:Heff} describe the exchange interaction between Fe ions and surrounding Cr and Fe ions and the interaction with an external magnetic field, respectively.

The effective operator of the interaction of the Fe$^{2+}$ ion with an electric field, which takes into account the mixing of states with opposite parity, i.e. $3d^6$ and $3d^5 4p$ as well as $3d^6 S^{2-}$ and $3d^7 S^{1-}$, where $S$ refers to the electronic shell of the sulfur ions, is written as follows \cite{Eremin2019B}:
\begin{eqnarray}
\label{eq:He}
H_{E}=&&\sum_{\substack{p, t\\k=2,4}}
\left\{E^{(1)} U^{(k)}\right\}_{t}^{(p)}\times \\
&&\times \sum_{j} d^{(1 k) p}\left(R_{j}\right)(-1)^{t} C_{-t}^{(p)}\left(\vartheta_{j},\varphi_{j}\right)\nonumber
\end{eqnarray}

The curly brackets denote the Kronecker product of the spherical tensor of the electric field
($E_0^{(1)}=E_z$, $E_{\pm1}^{(1)}=\mp(E_x \pm i E_y)/\sqrt{2}$) with the unit irreducible tensor operator acting on the $3d$ electronic states. The spherical coordinates $R_j$, $\vartheta_j$, $\varphi_j$ denote the positions of the lattice ions (as in crystal field theory), and $C_{t}^{(p)}(\vartheta_j,\varphi_j)=\sqrt{4\pi/(2p+1)}Y_{p t}(\vartheta_j,\varphi_j)$ are the components of the spherical tensors. The quantities $d^{(1k)p}(R_j)$ are calculated in the local coordinate system with the $c$-axis along the $3d$-ion ligand direction.  For iron ions in an undistorted tetrahedral environment $p=3$, and, therefore, there are only two intrinsic parameters \cite{Eremin2019B}: $d^{(12)3}$ and $d^{(14)3}$, which we will try to extract based on relative intensities of transitions in the absorption spectra (Note that $d^{(1k)p}$ includes a Lorentz local-field correction factor). 

The effective coupling operator of Fe$^{2+}$ with an ${}^5 E$-state in an applied electric field is then  written as \cite{Eremin2019}
\begin{equation}
\label{eq:H1}
    H_{1} = -\frac{4\lambda}{3\Delta}\sqrt{\frac{2}{21}}\Big\{ 2d^{(12)3} - \sqrt{\frac{5}{3}}d^{(14)3} \Big\} \mathbf{E}\mathbf{S} U_{\alpha}
\end{equation}

Considering the results of Mössbauer studies  in the orbitally ordered ground state for  $T<T_{\mathrm{OO}}$ the orbital degeneracy of the lowest-lying Fe states is lifted \cite{Spender1972}. The measured electric field gradient $V_{zz}$ has a negative sign and the associated asymmetry parameter is about $0.2$, i.e. that the ground state of the iron ion is mainly a spin multiplet ${}^5 E_{\theta}$.

Within the states of the ${}^5 E_{\theta}$ multiplet, the effective operator \eqref{eq:H1} is supplemented by the following expression
\begin{eqnarray}
\label{eq:H2}
    && H_{2} = -\frac{\lambda}{3\Delta}\sqrt{\frac{2}{7}}\Big\{ d^{(12)3} - \frac{1}{2}\sqrt{\frac{5}{3}} d^{(14)3} \Big\} |\theta\rangle\langle\theta|\times \\
    &&\times\left\{\begin{aligned}
       E_x \Big[ \mathbf{S}_{Fe} \times (\mathbf{S}_{\xi'}^{(1)} + \mathbf{S}_{\xi'}^{(2)} + \mathbf{S}_{\eta'}^{(3)} + \mathbf{S}_{\eta'}^{(4)}) \Big]_x \\
     -  E_y \Big[ \mathbf{S}_{Fe} \times (\mathbf{S}_{\xi'}^{(1)} + \mathbf{S}_{\xi'}^{(2)} + \mathbf{S}_{\eta'}^{(3)} + \mathbf{S}_{\eta'}^{(4)}) \Big]_y \\
    -   E_x \Big[ \mathbf{S}_{Fe} \times (\mathbf{S}_{\xi'}^{(1)} + \mathbf{S}_{\xi'}^{(2)} - \mathbf{S}_{\eta'}^{(3)} - \mathbf{S}_{\eta'}^{(4)}) \Big]_y \\
    +   E_y \Big[ \mathbf{S}_{Fe} \times (\mathbf{S}_{\xi'}^{(1)} + \mathbf{S}_{\xi'}^{(2)} - \mathbf{S}_{\eta'}^{(3)} - \mathbf{S}_{\eta'}^{(4)}) \Big]_x \\
        \end{aligned}\right\}\nonumber
\end{eqnarray}
Here we have used the notations $\mathbf{S}_{\xi'}^{(1)}=\sum_{Cr}(J^{(1)}_{\xi', Cr}/\Delta) \mathbf{S}_{Cr}$, $\mathbf{S}_{\eta'}^{(3)}=\sum_{Cr}(J^{(3)}_{\eta', Cr}/\Delta) \mathbf{S}_{Cr}$  etc.

\begin{figure}[]
\includegraphics[clip,width=7.5cm]{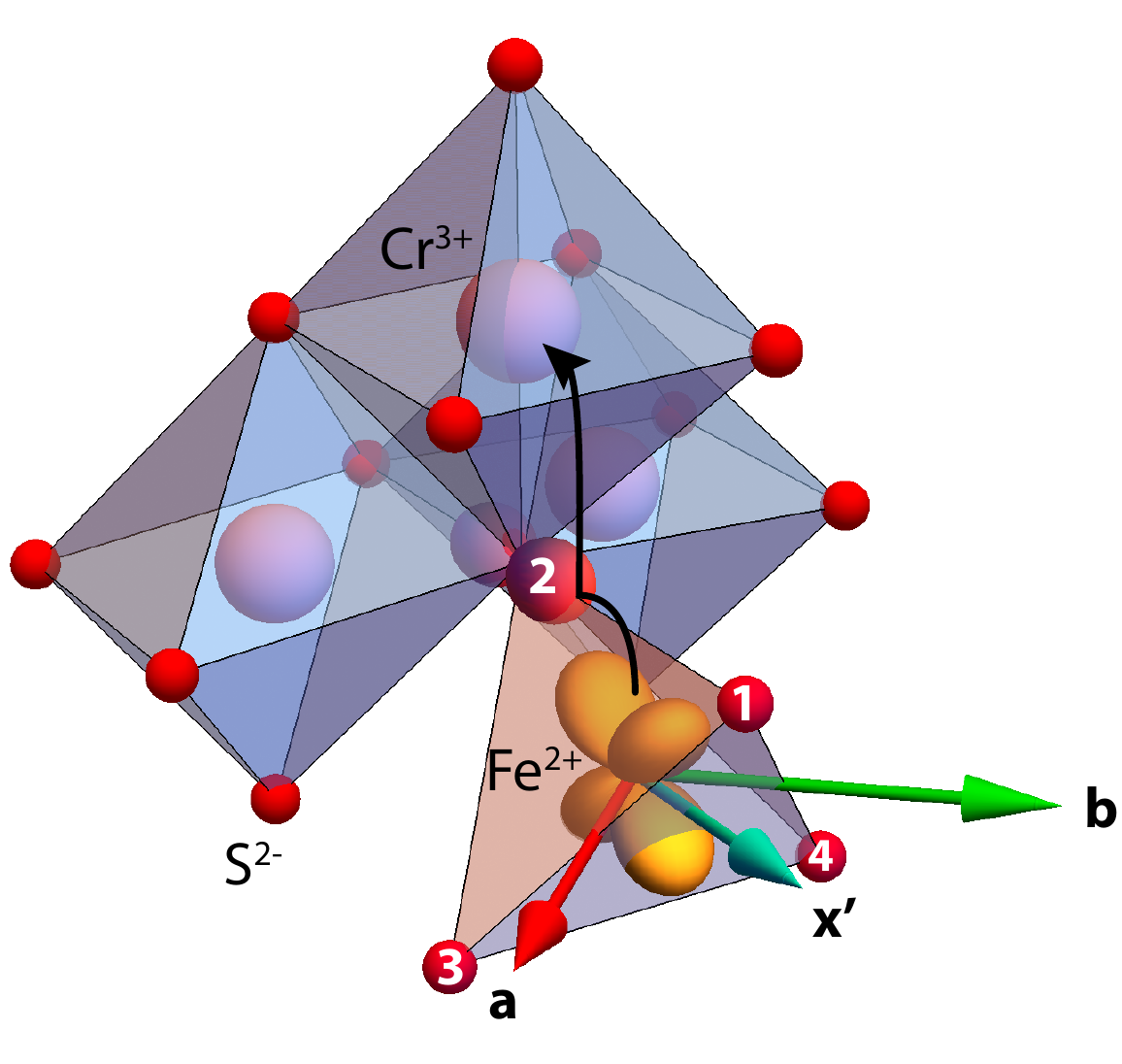}
\caption{\label{fig:fragment} Fragment of the FeCr$_2$S$_4$ crystal structure illustrating the dominant superexchange coupling (black arrow) of the Fe(1) excited state $\xi'$ to a neighbouring Cr ions via the $p$-state of the sulphur ligand with number 2.  }
\end{figure}

  The operator $H_2$ describing the coupling of Cr and Fe spins with an electric field \eqref{eq:H2} was obtained by combining Eq.\;\eqref{eq:He} with the operator of exchange and spin-orbital interactions in the third order of perturbation theory, similar to the approach described in Ref. \onlinecite{Eremin2019B}. The upper index at the parameters of the exchange interaction corresponds to the numbering of the sulfur ions in Fig. \ref{fig:fragment}, through which the superexchange interaction of the excited state of the Fe ions with the nearest three Cr ions is taking place. Variables with an upper prime refer to the local coordinate system with the $\mathbf{x}'$-axis directed parallel to the edge of the tetrahedron.

The effective operator $H_2$ contains a vector product of Fe and Cr spin operators and, therefore, turns to zero in the case of a collinear order of spins. Experimentally, the presence of non-collinearity between spin orientations in \fcs was found in Ref.~\onlinecite{Kalvius2010}.  The exact spin configuration and, thus, the canting angle between Fe and Cr spins angles is not yet known.

Therefore, we determine the magnitude and direction of the exchange field of the Cr ions based on the observed relative intensities and positions of the absorptions  $E_2-E_4$ in the spectrum. The direction of the Fe spin is calculated on the basis of the ground state wave function of the Hamiltonian \eqref{eq:Heff}, while the direction of the Cr spins is assumed to be coinciding with the direction of the effective exchange field. The angle between Cr and Fe spins obtained by this procedure is $170^{\circ}$.

Finally, there is another effective on-site interaction of the Fe spin with the electric field, which in our case is described by the following operator
\begin{eqnarray}
\label{eq:H3}
    H_3 = E_z \frac{\lambda^2}{\Delta^2}\frac{4}{3}\sqrt{\frac{7}{2}} d^{(12)3}
    (S_x S_y+S_y S_x) |\theta\rangle\langle\theta|.
\end{eqnarray}
This operator $H_3$ is also derived in the third order of perturbation theory, when we take into account the virtual excitation processes caused by the spin-orbital interaction and the action of an induced electric field within the excited  ${}^{5}T_{2}$-states of the Fe$^{2+}$ ions.

\subsection{Extracting $d^{(12)3}/d^{(14)3}$ ratio from optical conductivity spectrum}
\begin{figure}[t]
\includegraphics[clip,width=9cm]{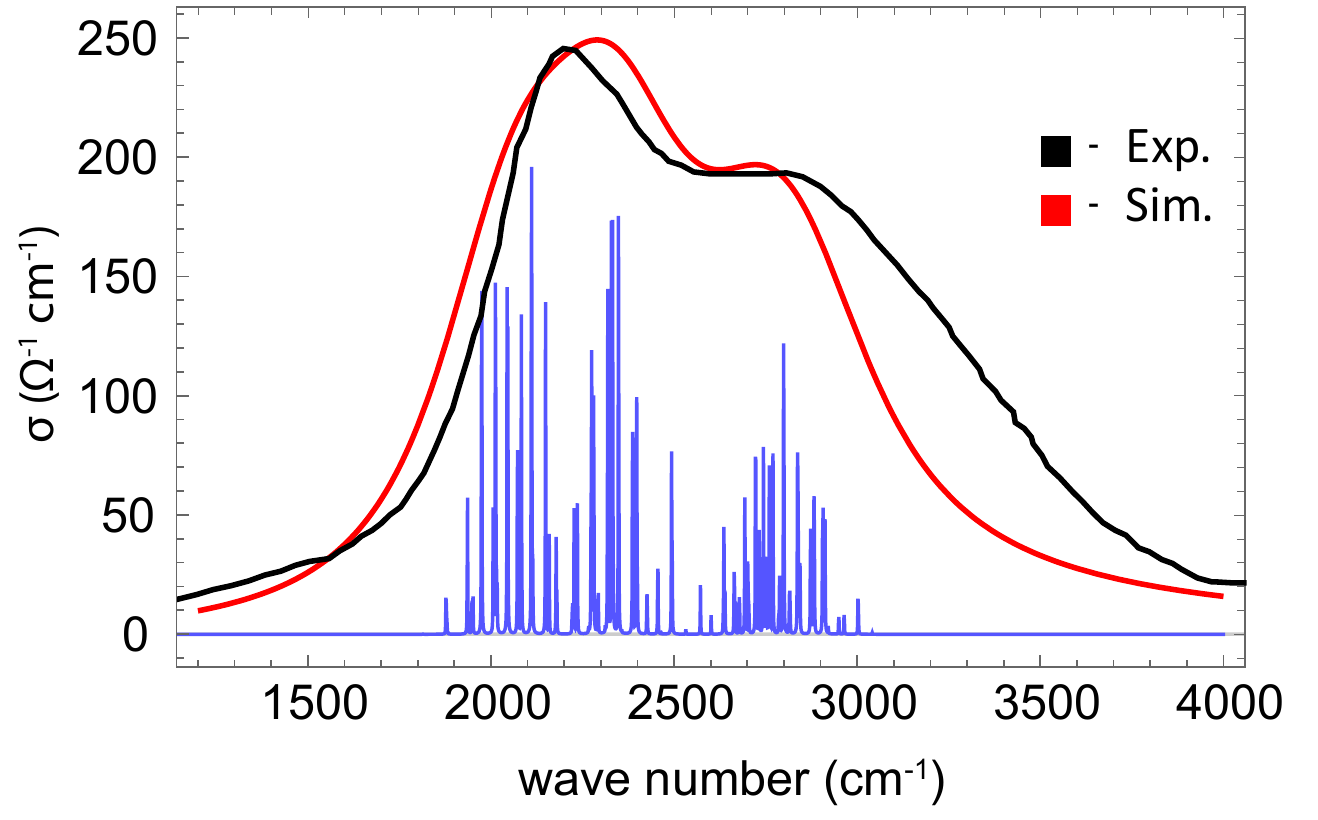}
\caption{\label{fig:opticalspec} Optical conductivity spectrum at $T=300K$: black curve – experimental data  taken from \cite{Ohgushi2008}, red cruve - our simulation using Eq.\;\eqref{eq:intens2}, blue curve - calculated superfine structure for $\gamma_j \sim 0$.}
\end{figure}

To gain more information on the parameters of the Hamiltonian $H_E$ given by Eq. \eqref{eq:He}, we start with the discussion of the optical $d-d$ transitions between the $^5 E$ and $^5 T$ multiplets, which were observed in Ref.~\onlinecite{Ohgushi2008} at $T=300K$. At this temperature the Fe$^{2+}$ tetrahedron is not distorted and there is no internal (molecular) exchange field due to the absence of orbital and magnetic ordering at room temperature. The splitting of the ground $^5 E$ multiplet due to the second order of spin-orbit coupling and spin-spin interaction is much smaller than the energy interval between $^5 E$ and $^5 T$ states and the splitting of the $^5 T$-multiplet is caused by the first order in spin-orbit coupling. According to \cite{Ohgushi2008} the optical conductivity spectrum can be approximated as follows
\begin{equation}
\label{eq:intens2}
    \sigma = \sum_j\frac{S_j \omega^2 \omega_j \gamma_j}{(\omega_j^2-\omega^2+\gamma_j^2)^2 + 4\omega^2 \gamma_j^2},
\end{equation}
where $\omega_j$ and $\gamma_j$ correspond to the eigenfrequencies and damping constants for the excitations between the ground multiplet and the $^5 T$-states split by the spin-orbit interaction with $\zeta \simeq 596$~cm$^{-1}$ \cite{Ohgushi2008}. We expanded the suggested model by taking into account a superfine structure caused by the mixing of ${}^5 E$ and ${}^5 T$ states due to spin-orbit coupling. This mixing leads to a violation of the Lande interval rule for exited states and actually increases the number of possible transition frequencies $\omega_j$. We calculated the transitions with wavefunctions in a basis of states $|{}^5 D, M_l, M_s\rangle$ by numerical diagonalization of the $25\times 25$ matrix. Using the pure operator of interaction with an electric field \eqref{eq:He} we expressed the oscillator strength $S_j$ using the two parameters $d^{(12)3}$ and $d^{(14)3}$
\begin{equation}
    S_j \propto \frac{1}{Z}(e^{-\frac{E_i}{k T}}-e^{-\frac{E_j}{k T}}) \omega_{ij} {|\langle j | H_E | i \rangle |}^2,
\end{equation}
where $Z$ is the partition function summing over all $25$ states. Our results are presented in Fig.~\ref{fig:opticalspec}. We assumed the same $\gamma_j \simeq 230$~$cm^{-1}$ for all oscillators. The obtained superfine structure is shown as blue spikes in Fig.~\ref{fig:opticalspec} denoting underlying transitions.

By varying the positions of the excitations we obtained the parameter  $\Delta\simeq 2300$~$ cm^{-1}$, which is somewhat smaller than the value reported in Ref.~\onlinecite{Ohgushi2008}. The relative intensities of the broadened lines allowed us to fix the ratio $d^{(12)3}/d^{(14)3} \sim 4$.

\subsection{Simulation of absorption spectra and estimation of the resulting polarization}

Having discussed the relevant coupling terms, we simulated the  absorption spectra corresponding to the presented Hamiltonians.

To reduce the number of unknown parameters, we considered $\lambda\simeq -70 cm^{-1}$ according to the estimates in Ref. \onlinecite{Laurita2015} for the related system FeSc$_2$S$_4$ and $J^{(1)}_{\xi', Cr}\simeq 17 cm^{-1}$ (see discussion in Ref. \onlinecite{Eremin2019}). The energy intervals between the electronic levels are mainly determined by the value of the exchange molecular field. The wave functions of the energy levels are labeled by the spin quantum numbers $(S,M_S)$. The spin-orbit interaction destroys the equidistant level spacing  and leads to the mixing of states with different quantum numbers $M_S$. In order to describe the transition from the ground state ($M_S=-2$)  to the excited state ($M_S=0$), which corresponds to the most intense spectral line $E_3$, the presence of a sufficiently large component of the exchange field at the site of the  Fe$^{2+}$ ions in the $ab$-plane must be assumed. 

\begin{figure}[t]
\includegraphics[clip,width=9cm]{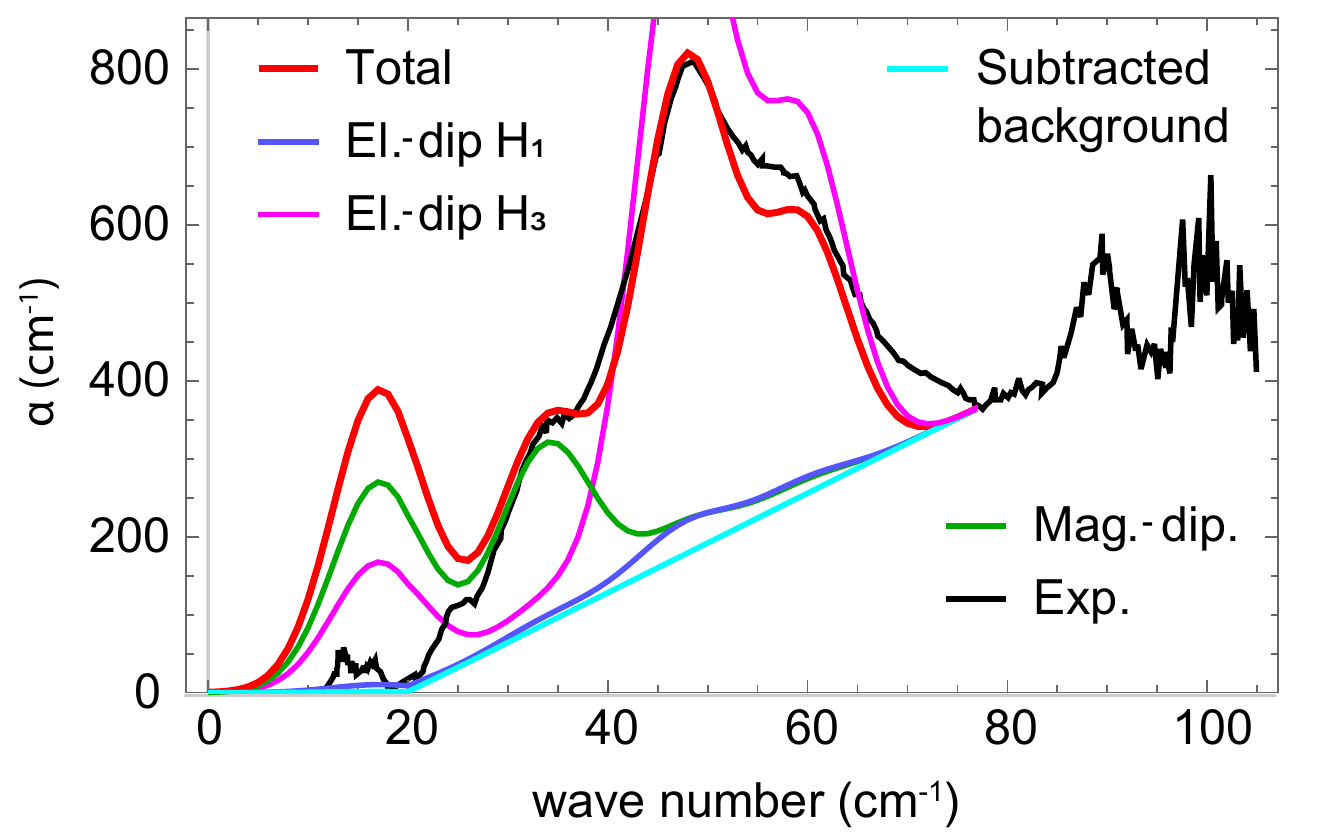}
\caption{\label{fig:sim} Absorption THz spectrum at $T=2K$. Black curve – experimental data, blue - calculation using Eq.\;\eqref{eq:H1}, cyan -  subtracted background, magenta - Eq.\;\eqref{eq:H3}, green - contribution of magnetic-dipole transitions, red - simultaneously taken into account magnetic and electro-dipole transitions.}
\end{figure}

The observed absorptions of the $E$-band where simulated as excitations from the ground state $|0\rangle$ to the excited states $|m\rangle$ using
\begin{eqnarray}\label{eq:intens}
    \alpha(\omega) \propto \sum_{m}|\langle 0| H_1 + H_2 && + H_3 + H_M |m \rangle |^2 \times\nonumber\\
    &&\times\,\omega^3 g(\omega_{0m} - \omega),
\end{eqnarray}
where $H_{M}=n g_s |\mu_b|\mathbf{S}\left[\frac{\mathbf{k}^{\omega}}{k}\times\mathbf{E}^{\omega}\right]$ describes the interaction of the effective spin $S=2$ with the magnetic field $\mathbf{H}^{\omega}$ of the incident THz radiation, giving rise to magnetic-dipole transitions. The refractive index $n$ sets the relation between $E^{\omega}$ and $H^{\omega}$ in Gaussian units. The sum $(H_1+H_2+H_3)$ represents the effective interaction between the spin and the electric field component $E^{\omega}$ of the radiation. Note that Boltzmann occupation factors were omitted in  Eq.~\eqref{eq:intens}, since the energies of the exited states of the effective Hamiltonian \eqref{eq:Heff} are all above 10 cm$^{-1}$. To simulate the experimental spectrum the shape function $g(\Delta\omega)$ \eqref{eq:intens} was treated as a Gaussian lineshape with the same width for all transitions. The calculated contributions to the absorbtion spectrum of the $E$-band in zero magnetic field is shown in Fig.~\ref{fig:sim}. A very strong interference between the different contributions to the total intensity of the absorption was observed. For example, considering only $H_3$ and neglecting the contributions of $H_1$ and $H_M$ in Eq.~\eqref{eq:intens} results in a strong increase in intensity in the high-frequency absorptions as demonstrated by the magenta curve in Fig.~\ref{fig:sim}. The intensity of transitions caused by $H_2$ is negligible compared to others, so it is not shown  separately. However, the operator \eqref{eq:H2} has a significant contribution to the $p_x$, $p_y$ components of the electric polarization which will be discussed below.

The background contribution to the absorption spectrum due to infrared active phonons and possible magnon modes was approximated by a straight cyan line.

The obtained set of parameters is $V\rho \simeq 150\;\mathrm{cm}^{-1}$, $\phi \simeq 11^{\circ}$, $J_{Fe,j}\langle \mathbf{S}_j\rangle \simeq 12.8\; \mathrm{cm}^{-1}$, $d^{(12)3} \simeq -5.563\; a.u$.

The set of angles determining the equivalent directions of the exchange field are given in Table \ref{tab:sets} as spherical angles $\vartheta$, $\varphi$. The contribution due to electric-dipole transitions determined by Eq.\;\eqref{eq:H3} dominates the spectrum.



\begin{table}[h]
\squeezetable
\begin{ruledtabular}
\centering\footnotesize
   \begin{tabular}{cccccccc}
   $\vartheta$ & $\varphi$ & $\langle S_a \rangle$ &  $\langle S_b \rangle$ & $\langle S_c \rangle$ & $\langle p_a \rangle$ &  $\langle p_b \rangle$ & $\langle p_c \rangle$\\
    \midrule
	 $106.5^{\circ}$ & $4^{\circ}$ & -1.85 & -0.17 & 0.21 & 0.20 & 2.60 & -3.77\\
	 $106.5^{\circ}$ & $176^{\circ}$ & 1.85 & -0.17 & 0.21 & 0.20 & -2.60 & 3.77\\
	 $106.5^{\circ}$ & $184^{\circ}$ & 1.85 & 0.17 & 0.21 & -0.20 & -2.60 & -3.77\\
	 $106.5^{\circ}$ & $356^{\circ}$ & -1.85 & 0.17 & 0.21 & -0.20 & 2.60 & 3.77\\
	
	 $73.5^{\circ}$ & $4^{\circ}$ & -1.85 & -0.17 & -0.21 & -0.20 & -2.60 & -3.77\\
	 $73.5^{\circ}$ & $176^{\circ}$ & 1.85 & -0.17 & -0.21 & -0.20 & 2.60 & 3.77\\
	 $73.5^{\circ}$ & $184^{\circ}$ & 1.85 & 0.17 & -0.21 & 0.20 & 2.60 & -3.77\\
	 $73.5^{\circ}$ & $356^{\circ}$ & -1.85 & 0.17 & -0.21 & 0.20 & -2.60 & 3.77\\
    \end{tabular}
    \caption{\label{tab:sets} Calculated expectation values for the spin components of the Fe ions and for the electric polarization vector $\bf p$ (in $10^{-3} a.u.$) per one Fe site. The angles refer to the exchange field acting on Fe(1). The projections of spin and polarization onto the $c$-axis are the same for the Fe(2) position, while the $a$ and $b$ components change sign.}
\end{ruledtabular}
\end{table}


Now we turn to the discussion of the change in the absorption spectrum in the external magnetic field. Consequently, the intensities of electric dipole transitions defined by expressions \eqref{eq:H1} and \eqref{eq:H3} are roughly speaking proportional to the square of the dielectric permittivity.  According to Fig. \ref{fig:deltaEps}, when the magnetic field increases for $E^{\omega}\parallel H$, the dielectric constant $\varepsilon'\sim n^2$ increases and decreases for $E^{\omega}\perp H$. Qualitatively, this corresponds to the trend of changes in the absorption intensity in Figs. \ref{fig:Voigtspectra}(b)-(d) and agrees with the conclusion that the absorption lines $E_2$, $E_3$ and $E_4$ are mainly due to the electric component of the electromagnetic wave.
\begin{figure}[]
\includegraphics[clip,width=9cm]{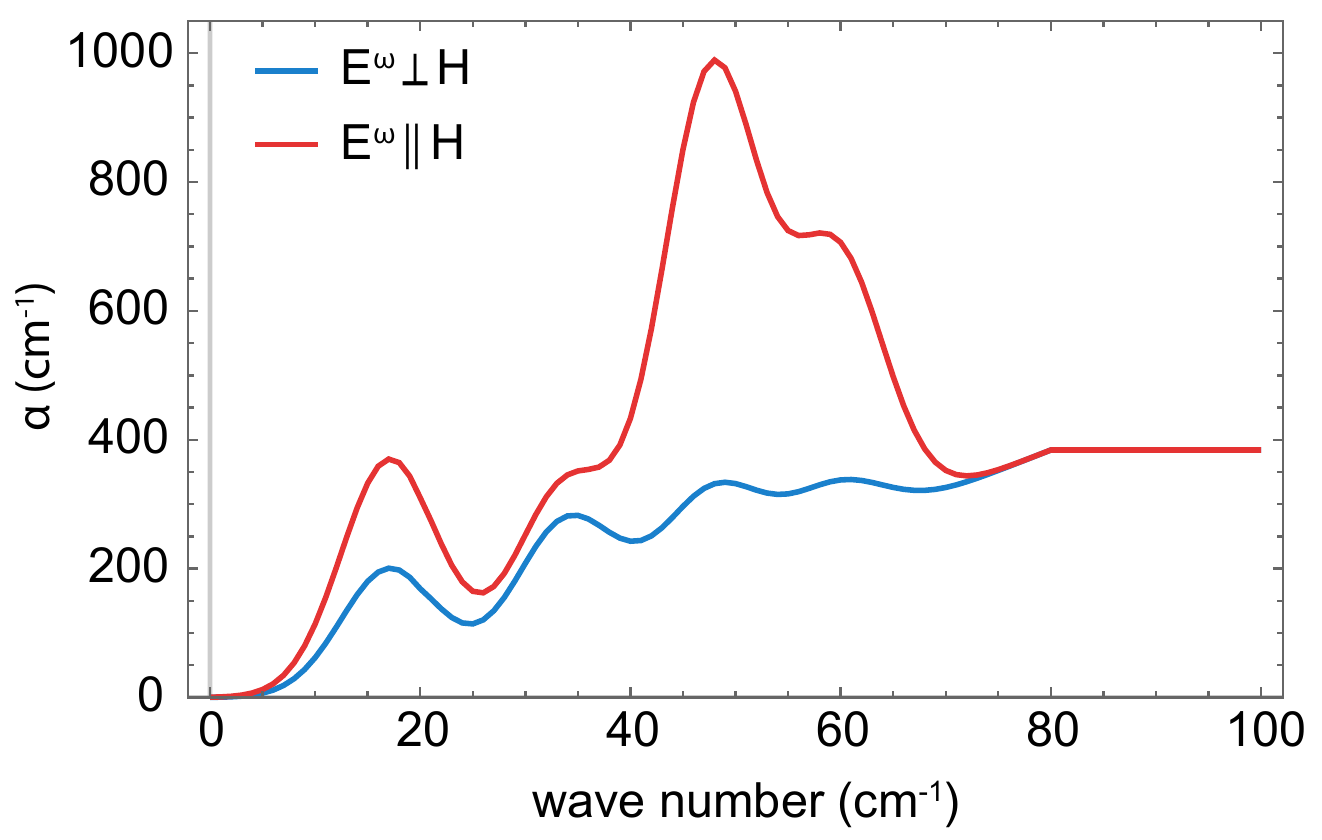}
\caption{\label{fig:simvf} Calculated THz absorption spectrum at $2K$ for domains, aligned along the $\bf c$-axis. Blue curve – Faraday geometry: $H^\omega \parallel \mathbf{c}$, $E^\omega \perp \mathbf{c}$, red – Voigt configuration:  $E^\omega \parallel \mathbf{c}$,  $H^\omega \perp \mathbf{c}$.}
\end{figure}

Passing on to the description of the microscopic theory of changes in the absorption spectrum when an external magnetic field is applied, we note the following. From the experimental spectra we can assume that the absorption spectrum of the $E$-band almost does not change its shape in an applied magnetic field. Moreover, we determined with our simulation that the line positions and relative intensities are mainly determined by the internal molecular field.


We believe that the external magnetic field makes preference to domains where the  $c$-axis is parallel to  the applied field. The calculated spectra for such domains are shown in Fig. \ref{fig:simvf}. The tendency of the change in absorption in applied external fields is in agreement with experiment.


It is important to emphasize that our model provides a consistent description of reported Mössbauer data \cite{Spender1972} regarding the asymmetry parameter ($\eta = 0.23$) and the sign of the electric field gradient at the Fe nucleus ($V_{zz}<0$), which are mainly determined by $V$ and $\phi$. It is also interesting to note the following. The minimum energy of the operator \eqref{eq:Heff} depends on the direction of the exchange (molecular) field acting on iron spin. By adjusting the magnitude and direction of the exchange field according to the observed absorption spectrum, we were able to calculate the magnitude and direction of the iron spins as a result of the diagonalization of the Hamiltonian \eqref{eq:Heff}. Performing such kind of calculations, we found that the angle between the directions of the molecular field, which is presumably determined by the total direction of chromium spins, and the Fe spins is about $\sim 170^{\circ}$. This is an interesting mechanism for the formation of a non-collinear spin arrangement of chromium and iron spins, which to the best of our knowledge, has not been reported before.



As one can see from Table \ref{tab:sets}, in the absence of an external magnetic (electric) field, there are eight energetically equivalent spin configurations differing in relative orientation of iron, chromium spins and spontaneous electric polarization. The absolute values of the spin-induced electric polarization components along the \textbf{c}-axis are equal, but they differ in sign. Therefore, we can speculate about two different types of electrically polarized domains with opposite electric polarization in \fcs. When an external electric (magnetic) field is switched on, the equivalence of these domains gets broken. In this regard, we can understand why the evaluated electric polarization along the $c$-axis for the monodomain case (after averaging over Fe(1)S$_4$ and Fe(2)S$_4$ fragments) is about $P=255\mu C/m^2$, i.e. it is larger by the factor $\sim 3.6$ than the value reported in Ref.~\onlinecite{Lin2014}.

In addition, we want to point out that the reported symmetry lowering with a loss of inversion symmetry at the transition to the orbitally ordered state $T<9$~K as reported in Ref. \onlinecite{Deisenhofer2019} can, in addition, lead to contributions of the electric polarization considered here, because the presence a spontaneous internal electric field will enhance the non-equivalence of different domains.

Moreover, we want to comment on the fact, that the obtained value $J_{Fe,j}\langle \mathbf{S}_j\rangle \simeq 12.8$~$cm^{-1}$ is rather small. Comparing this value with the exchange field of chromium spins $12 J^{(1)}_{\xi', Cr}\langle \mathbf{S}_{Cr} \rangle$ one can conclude that the average projection of Cr spins $\langle \mathbf{S}_{Cr} \rangle$ along the direction of the Fe spin, perhaps, is reduced due to the non-collinearity between Cr$^{3+}$ spins. These issues obviously require further investigation.

\section{Summary}

We identified six low-frequency modes in the multiferroic ground state of \fcs by THz-spectroscopy and studied their behavior in magnetic fields up to 7~T. The intensity dependence of the three most intense modes $E_2-E_4$ on the relative orientation of the light polarization and the external magnetic field allowed to conclude that they are predominantly electric-dipole active. Modes $M_1$ and $E_{1b}$ are active for $E^\omega \perp H$, while for $M_2$ and $E_{1a}$ no clear selection rules could be determined.
In addition, a theoretical model is introduced to describe the excitations $E_1-E_4$ in terms of the low-energy electronic excitations of the Fe$^{2+}$-ions ($3d^6$, $S\,=2$) in an tetrahedral S$^{2-}$ environment. Reproducing the  eigenfrequencies and relative intensities of these absorption lines gives a good agreement for the strongly field-dependent modes $E_2-E_4$, but overestimates the intensity of mode $E_1$. The obtained parameters and effective Hamiltonians also allow to reproduce experimental parameters of previous Mössbauer studies and the order of magnitude of the electric polarisation induced by orbital ordering and non-collinear spin ordering. The additionally observed modes $M_1$ and $M_2$ are not described within our theoretical approach. Hence, further theoretical and experimental work on single crystals will be needed to decide, whether they correspond to collective magneto-electric magnon modes of the ground state.

\begin{acknowledgments}
We acknowledge support by the Deutsche Forschungsgemeinschaft via TRR 80 (project no. 107745057).
The work of M.V.E. and K.V.V. was supported by the Russian Science Foundation (Project No. 19-12-00244).
\end{acknowledgments}

\end{document}